# Discovery of room temperature ferromagnetism in metal-free organic semiconductors

Qinglin Jiang[1†], Jiang Zhang[1,2†*], Zhongquan Mao[2†], Yao Yao[1,2*], Duokai Zhao[1], Wenqiang Zhang[1], Jiadong Zhou[1], Nan Zheng[1], Huanhuan Zhang[1], Manlin Zhao[1], Yong Wang[3], Xiaolong Li[3], Dehua Hu[1*] & Yuguang Ma[1*].

[1] State Key Laboratory of Luminescent Materials and Devices, South China University of Technology, Guangzhou 510640, China.

[2] Department of Physics, South China University of Technology, Guangzhou 510640, China.

[3] Shanghai Institute of Applied Physics, Chinese Academy of Sciences, Shanghai Synchrotron Radiation Facility, Shanghai 201800, China.

*Correspondence to: jonney@scut.edu.cn (J.Z.), yaoyao2016@scut.edu.cn (Y.Y.), msdhhu@scut.edu.cn (D.H.H.) and ygma@scut.edu.cn (Y.G.M.).

†These authors contributed equally to this work.

Creating magnetic semiconductors that work at warm circumstance is still a great challenge in the physical sciences. Here, we report the discovery of ferromagnetism in the metal-free perylene diimide semiconductor, whose Curie temperature is higher than 400 Kelvin. A solvothermal approach is used to reduce and dissolve the rigid-backbone perylene diimide crystallites, and radical anion aggregates were fabricated by the subsequent self-assembly and oxidation process. Magnetic measurements exhibit the ferromagnetic ordering with the saturated magnetization of 0.48 $\mu_B$ per molecule and the appreciable magnetic anisotropy. X-ray magnetic circular dichroism spectra suggest the ferromagnetism stems from π orbitals of radicals. Our findings unambitiously demonstrate the long-range ferromagnetic ordering can survive at room temperature in organic semiconductors, although which are intuitively regarded to be nonmagnetic.

**Main Text:**

In the early scenario of quantum mechanics, the ferromagnetism must arise from the electrons with the principal quantum number being greater than two *(1)*. Since the 1990s, chemists have been persistently exerting efforts to synthesize organic radical ferromagnets carrying only s and p orbitals *(2-4)*, and thus far the spontaneous magnetization of such materials was solely observed below 36 Kelvin *(5, 6)*. In order to improve the Curie temperature, heavy atom radicals *(7, 8)* and insulating metal-organic coordination complexes *(9-15)* were introduced. However, room temperature ferromagnetism, along with semiconducting properties in the organic materials composed merely of carbon, hydrogen, oxygen, and nitrogen have hitherto not been found. Due to the negligible spin-orbital couplings and thus the long spin lifetime in organic materials, the room temperature ferromagnetism appeared in organic semiconductors will definitely pave a new way to the realization and application of room temperature magnetic semiconductors.

As a widely-investigated n-type semiconductor, perylene diimide (PDI) serves as a promising candidate in various optoelectronic devices including thin-film transistors, photovoltaics and light-emitting diodes *(16-18)*. The PDI molecule comprises a perylene backbone with two imide end groups. Due to the strong intermolecular π-π interaction and hydrogen bonding, the pristine PDI tends to crystallize and suffers poor solubility in commonly-used organic solvents, impeding the manufacturing of solution processable organic thin film devices *(19)*. With four electron-withdrawing carbonyl groups, the PDI molecule can be reduced to metastable radical anions or dianions through electrochemical or chemical reduction *(20, 21)*.

We conducted a reduction and dissolution treatment of the sublimated PDI crystallites with a π-π stacking distance of 3.34 Å (Fig. S1) by the solvothermal method. A purple suspension was obtained after cooling down, which was subsequently used to prepare a film by drop casting in the glove box ($H_2O$ & $O_2$ < 1 ppm). The grazing incidence wide-angle X-ray scattering (GIWAXS) image exhibits visible arcs of diffracted signal (Fig. 1A), implying the formation of crystallite with edge-on orientation. The line-cut profiles manifest a sharp (011) diffraction in the out-of-plane direction at 0.73 Å$^{-1}$ and ($12\bar{2}$) diffraction in the in-plane direction at 1.89 Å$^{-1}$ (Fig. 1B). The ($12\bar{2}$) diffraction peak is identified as the π-π stacking, and the calculated distance is 3.32 Å, slightly shorter than that of the sublimated PDI crystallites (3.34 Å). X-ray diffraction also corroborated the π-π stacking spacing was shortened via the unique self-assembling approach (Fig. 1C). The scanning electronic microscopy (SEM) photograph of the films displays compactly stacked self-assembly nanorods of 50-60 nm in width aligning on the substrate (Fig. 1D). It is essential that the π-π stacking keeps close after the reduction process considering the PDI species are ionized.

After exposed in the ambient air, the color of the PDI films quickly changed from olive to dark gold implying the species are changed. Benefitting from the intrinsic semiconducting features, the UV-vis absorption spectra were employed to measure the valence states. The main absorption peak at 554 nm indicates the PDI dianion species dominate the suspension and the as-prepared film (Fig. 1E). During the fast oxidation process, the appeared peaks at 773, 810 and 965 nm figure out the production of the PDI radical anions (Fig. 1E). The electron paramagnetic resonance (EPR) spectra were adopted to probe the formation of the radical anions in the PDI films after exposed to air. A strong resonance signal around g = 2.003 was observed, which originates from unpaired electrons, clearly indicating the dramatic increase of radical anion after oxidation (Fig. 1F). It is worth mentioning that, different from the previous strategies that the radicals are obtained by a reduction process *(20, 21)*, our self-doping approach of radical anions is conducted with a spontaneous oxidation process from closed-shell dianions. Consequently, the self-assembly PDI aggregates synthesized by the present approach are unique in two aspects: the close π-π stacking distance and the large concentration of radical anions, which may introduce a strong spin exchange interaction between radicals and make the ferromagnetic ordering eventually emerge.

We investigated the magnetic properties of the PDI semiconductors with vibrating sample magnetometer. Magnetization-magnetic field (*M-H*) curves recorded at 10 K and 300 K show representative ferromagnetic hysteresis loops (Fig. 2A). For a comparison, linear *M-H* curve of silicon substrate, PDI raw material, and PDI film without exposure to air merely show diamagnetic features (Fig. S2). It strongly suggests the ferromagnetic signals arise from the oxidation of PDI film. The evolution of the magnetic properties from diamagnetic to ferromagnetic after oxidation indicates more definitely that the observed ferromagnetism are intrinsically derived from PDI film. The amount of the magnetic metallic impurities in our sample measured by particle-induced X-ray emission (PIXE) spectra is far from sufficient to induce the ferromagnetic signal (Table S1). The magnetization is saturated at 2 kOe, and the coercive field reaches 46 Oe at 300 K. The extracted saturated magnetization at 300 K is ~6.9 emu·g$^{-1}$, equivalent to ~0.48 $\mu_B$ per molecule, which is unexpectedly large in organic materials and far beyond the value in 36 K canting ferromagnets *(6)*, as well as recently reported weak ferromagnets in amorphous polymerized tetracyanoquinodimethane (TCNQ) *(22)* and 1,3,5-trizaine-linked porous organic radical frameworks *(23)*. Figure 2B displays the in-plane and out-of-plane M-H curves of the PDI film at 300 K. The distinct in-plane hysteresis loop and the higher out-of-plane saturation field clearly exhibit an anisotropic characteristic of ferromagnetism at room temperature. It indicates in-plane spin alignment and a magnetic easy axis in the plane of the film, i.e. along the long axis direction of the nanorods (Fig. 1C). Furthermore, we measured the hysteresis loops within the temperature range from 10 K to 400 K, and the temperature dependence of the coercive field (*H*c) also suggests the ferromagnetism above room temperature (Fig. 2C). The temperature dependence of magnetization in zero-field-cooled (ZFC) and field-cooled (FC) condition at the magnetic field of 20 Oe is displayed in Fig. 2D. The bifurcation between ZFC and FC plots starts from 400 K, indicating the Curie temperature of the ferromagnetic PDI film is above 400 K.

We employed the soft X-ray magnetic circular dichroism (XMCD) measurement to reveal the contribution of the elements to the observed ferromagnetism *(24-26)*. Figure 3 depicts the room temperature X-ray near-edge absorption spectroscopy (XANES) and XMCD results at the carbon K edge with the applied magnetic field 2 kOe along the X-ray beam direction, incident at a 15° angle on the ferromagnetic PDI film. The edge structure exhibit four prominent resonance peaks, which associate with the transition from the C 1s levels to the π* orbitals *(27)*. In Fig. 3B, the first peak at 284.2 eV stems from the transition from the C 1s levels of the perylene core C atoms to the lowest unoccupied molecular orbital (LUMO) (L0). The second peak at 285.8 eV belongs to the transitions from core levels to the higher excited energy orbital L1-L3. The peaks around 287.4 and 288 eV can be assigned to the transition from C atoms on imide moiety to L0 and L1-L3, respectively.

The XMCD signals were acquired by the difference in the normalized X-ray absorption spectra recorded with right- and left-circularly polarized beam. The remarkable XMCD signal of negative dichroism can be observed in the first peak at 284.2 eV and the second peak at 285.8 eV possesses positive dichroism under the applied magnetic field of 2 kOe (Fig. 3B), while it disappears in the non-ferromagnetic PDI raw material (Fig. 3C). Excitations to σ* orbitals, which should appear above 290 eV, display wider peaks and no visible magnetic signal. The room temperature XMCD signal proves that the first unoccupied states at L0 behave as magnetically ordered and the local ferromagnetism originates from π-conjugated electrons of the PDI film. We also collected the XANES spectra at inclined incidence 60° (Fig. S3). The intensity ratios for the π*-derived resonances peaks show a distinct change with incidence angle and peaks around 287.4 and

288 eV are more pronounced in XMCD spectrum, which indicates the molecular plane is nearly normal to the substrate surface *(27)*, and this coincides with the GIWAXS results. We measured the spectra at the Fe, Co, Ni, and Mn L-edge resonance, which do not appear detectable absorption resonance signals, further demonstrating the very few metallic impurities do not contribute to the magnetic orders. It is worth noting that the carbon K edge XMCD can only detect the contribution of orbital moments *(28)*, and the highly ordered polarized π orbitals in the rigid PDI molecules induce the observed anisotropy of the magnetization. As a consequence, the room temperature ferromagnetism in this organic material can be corroborated by various complementary magnetic probes.

The ferromagnetism could be persistently reserved when the film was returned to keep in the glove box without oxygen and water. We also found that after about three months the ferromagnetism changed to diamagnetism when the film was continuously exposed to air (Fig. S4). These phenomena suggest the ferromagnetism stems from the metastable radical anions which are generated during the oxidation process in air. We thus illustrate the formation process of the metastable radical anions in our sample as follows. The raw material is in the neutral state, and it is thoroughly reduced to the dianion state with high reduction potential by our solvothermal approach. Both states are of closed-shell structure and show the diamagnetism. We adopt the electrochemical method to determine the redox potentials of the dianion, the radical anion, and the neutral state (Fig. S5). In this energy-descending configurations, the sample will be spontaneously oxidized in air to produce the neutral state through an intermediate radical anion state with open-shell structure (Fig. 4A), which affords unpaired electrons for the emergence of the ferromagnetism. Interestingly, the production of the neutral state activates the second reaction path to generate the radical anions, namely the electron could transfer from the dianion state to the neutral state if the molecules are sufficiently close to each other (Fig. 4B). The produced radical anions from the second reaction are initially of singlet spin configuration since the reactants are of closed-shell structure. With the effect of zero-field splitting *(29)*, the triplet is generated through a spin flip process and it is inhibited to converse back to the neutral and dianion state due to the spin-forbidden transition. In a short duration, the detailed balance of three components is reached and sufficient metastable radical anions are formed through such a reaction. This might explain the long-standing ferromagnetism of the sample kept in the ambient air.

Room temperature ferromagnetism is rarely found in semiconductors *(30)*. As an imide-fused polycyclic hydrocarbon, the PDI film is naturally an n-type semiconductor (Fig. S6). According to Néel's local molecular field theory *(31)*, when the average radius of the open shell is roughly half of the interatomic distance, the exchange integral turns out to be positive giving rise to ferromagnetic exchange interactions. Through our unique synthesis approach, the PDI molecules are assembled into nanorods and the π-π stacking keeps close, enabling sufficiently strong ferromagnetic spin exchange interactions among radicals *(29, 32)* (Fig. S7). In this circumstance, the emergence of spontaneous magnetization can be illustrated with two possible theoretical models based upon localized and itinerant (delocalized) electrons. The first one was initially proposed by Heisenberg1, which refers to the direct or indirect exchange of local magnetic moments, and the second one was from Stoner's picture *(33, 34)* in which the energy band is spontaneously spin-split. In π-conjugated organic crystallites with good molecular packing, the electrons are regarded to be delocalized to form band-like transport channels, but due to the strong vibronic couplings, the electrons are also easy to be localized *(35)*. This dual localized and delocalized features of unpaired electrons in radical anions cooperatively give rise to the long-range ferromagnetic ordering in PDI semiconductors.

In conclusion, organic semiconductors, especially open-shell polycyclic hydrocarbons with metastable radicals and close stacking structure, are demonstrated as a significant candidate for the room temperature ferromagnets. We anticipate this novel material will be useful for investigating fundamental spin behaviors in the organic semiconductors, opening the door to explore the applications such as the pure-organic spin devices.

**Acknowledgments**

**Funding:** The work at the State Key Laboratory of Luminescent Materials and Devices was supported by the Natural Science Foundation of China (No. 21334002, 51521002, and 51403063), the Major Science and Technology Project of Guangdong Province (No. 2015B090913002), and the Foundation of Guangzhou Science and Technology Project (No. 201504010012). The work at the Department of Physics was supported by the National Natural Science Foundation of China (No. 21473211, 11574052, and 91833305) and the Key Technologies R&D Program of Guangzhou City (No. 201704020182 and 201803030008). Y.W. and X.L.L. acknowledge the support from the National Natural Science Foundation of China (No.11875314 and 11875315).

**Author contributions:** Y.G.M. supervised the project. J.Z., Q.L.J. and Z.Q.M. conceived the experiments. Q.L.J., D.H.H. and D.K.Z. performed the synthesis of the samples and the structure characterization with the assistance by W.Q.Z., J.D.Z., N.Z., H.H.Z. and M.L.Z.. J.Z. and Z.Q.M. conducted the measurements of the magnetic and transport properties. J.Z., Q.L.J, Z.Q.M. and Y.Y. performed the XMCD experiments with the assistance by Y.W. and X.L.L.. J.Z., Z.Q.M. and Y.Y. analyzed the data of the magnetic and XMCD measurements and figures preparation. Y.Y. proposed the model to explain the experimental results. J.Z., Q.L.J., Z.Q.M., Y.Y. and D.H.H. wrote the manuscript with the contributions from all authors. All authors discussed the results and commented on the manuscript.

**Competing interests:** The authors declare that they have no competing financial interests.

**Data and materials availability:** The data that support the findings of this study are available from the corresponding author on reasonable request.

**Supplementary Materials:**

Materials and Methods

Figures S1-S7

Tables S1

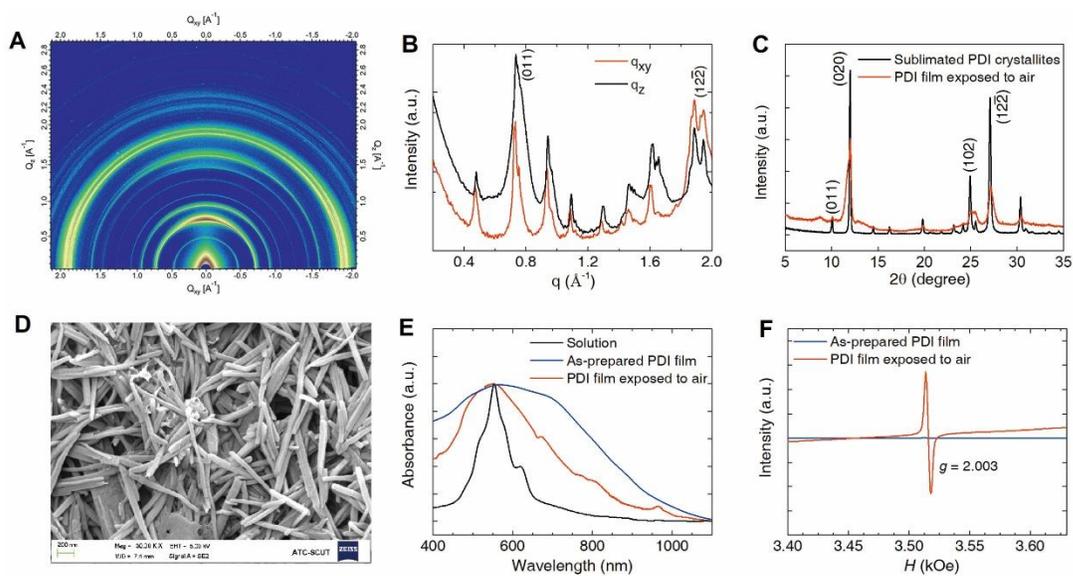

**Fig. 1. Structure characterization.** (**A**) Two-dimensional GIWAXS pattern of PDI film. (**B**) Out-of-plane and in-plane line-cut profiles of the film. The peaks at 0.73 Å$^{-1}$ and 1.89 Å$^{-1}$ correspond to (011) refraction in the out-of-plane direction and (12$\bar{2}$) refraction in the in-plane direction, respectively. (**C**) XRD patterns of the sublimated PDI crystallites and PDI film exposed to air. The (12$\bar{2}$) diffraction peak is identified as the π-π stacking. The interplanar distance in PDI film is slightly shorter than that of the sublimated PDI crystallites. (**D**) SEM image of PDI nanorods with width of 50-60 nm aligned on the silicon substrate. (**E**) UV-vis absorption spectra of the purple solution, the as-prepared PDI film, and the PDI film after exposed in ambient air. (**F**) EPR spectra of the as-prepared PDI film and the film exposed to air. A strong resonance signal around 3.52 kOe was observed in the PDI film exposed to air.

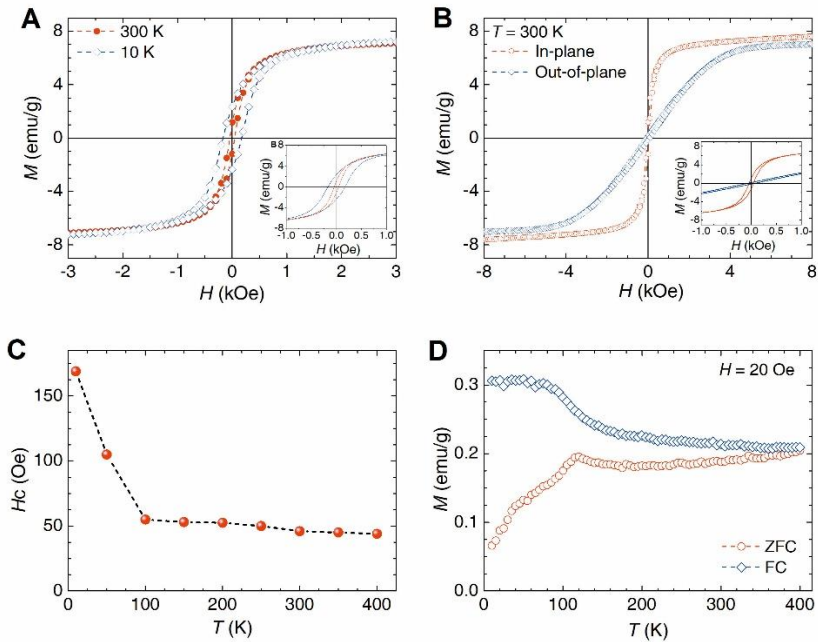

**Fig. 2. Magnetic properties of PDI film.** (**A**) The *M-H* hysteresis loops of PDI film taken at 10 and 300 K. The inset loops exhibit the coercive field of 46 Oe at 300 K and 169 Oe at 10 K. (**B**) In-plane and out-of-plane *M-H* hysteresis loops at 300 K. The inset shows the loops of the low field region. Higher out-of-plane saturation field demonstrates an anisotropic characteristics of ferromagnetism, indicating the magnetic easy axis is along the plane of the film. (**C**) Temperature dependence of coercive fields, suggesting the ferromagnetism above room temperature. (**D**) ZFC and FC magnetization curves with temperature measured at the applied magnetic field of 20 Oe. The bifurcation between ZFC and FC curves implies the Curie temperature of the PDI film is beyond 400 K.

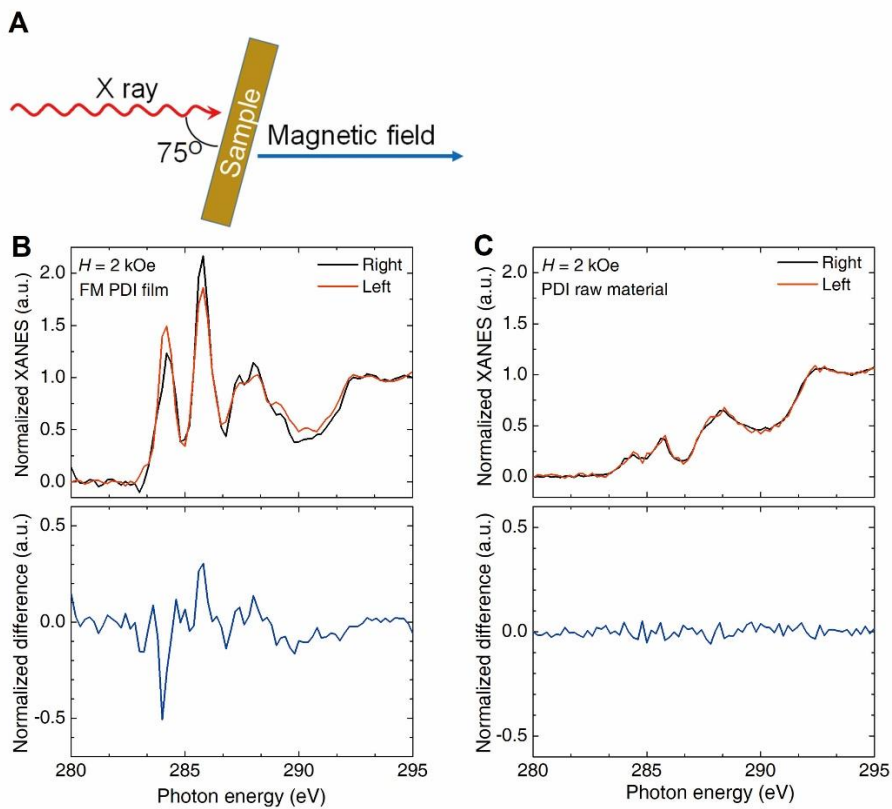

**Fig. 3. X-ray absorption of ferromagnetic PDI film and PDI raw material.** (**A**) Top view of the XMCD experiment geometry. (**B**) Normalized XANES spectra for the right (black) and left (red) photon helicity in the energy range of carbon K absorption edge recorded at 300 K with the applied magnetic field of 2 kOe for ferromagnetic (FM) PDI film. Underneath is shown the XMCD spectrum, defined as the absorption difference between right- and left-circularly polarized X-rays. The distinct XMCD signal of negative dichroism in the first peak at 284.2 eV and the second peak at 285.8 eV possesses positive dichroism were observed. (**C**) Normalized XANES spectra and difference spectra of polarization dependent absorption spectra indicate the disappearance of magnetic dichroism in the non-ferromagnetic PDI raw material.

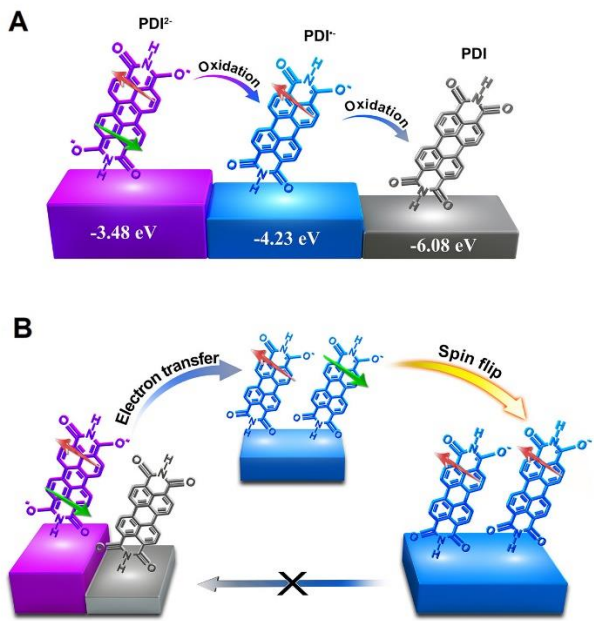

**Fig. 4. Formation and Stabilization of PDI·⁻.** (**A**) Energy levels of PDI films measured by electrochemistry. The highest occupied molecular orbital (HOMO) of PDI dianions (PDI$^{2-}$) is determined to be -3.48 eV, higher than the singly occupied molecular orbital (SOMO) of radical anion (PDI·⁻) (-4.23 eV) and the HOMO of PDI neutral state (-6.08 eV) (Fig. S5). (**B**) The PDI$^{2-}$ reacts with the PDI to produce two PDI·⁻ with singlet spin configuration in the first step. The singlet is then conversed to the triplet during spin flip processes, and the triplet is inhibited to converse back to the PDI and PDI$^{2-}$ due to the spin-forbidden transition.

# Supplementary Materials for

## Discovery of room temperature ferromagnetism in metal-free organic semiconductors

Qinglin Jiang[1†], Jiang Zhang[1,2†*], Zhongquan Mao[2†], Yao Yao[1,2*], Duokai Zhao[1], Wenqiang Zhang[1], Jiadong Zhou[1], Nan Zheng[1], Huanhuan Zhang[1], Manlin Zhao[1], Yong Wang[3], Xiaolong Li[3], Dehua Hu[1*], & Yuguang Ma[1*]

[*]Correspondence to: jonney@scut.edu.cn (J.Z.), yaoyao2016@scut.edu.cn (Y.Y.), msdhhu@scut.edu.cn (D.H.H.) and ygma@scut.edu.cn (Y.G.M.)
[†]These authors contributed equally to this work.

**This PDF file includes:**

Materials and Methods
Supplementary Text
Figs. S1 to S7
Tables S1

## Materials and Methods

Sample preparation

PDI powders (98%) were purchased from J & K Co., Ltd. Then, we used sublimation technology to obtain pure PDI powder. PDI powders (20 mg) were sealed in an autoclave and heated at 140 °C for 24 h. The autoclave containing the dianion solution was opened in a $N_2$ glovebox ($H_2O$ & $O_2$ < 1 ppm). High-resistance 4×4 mm silicon wafer with (111) orientation was subsequently cleaned with sonication in water, cleaning agent, acetone and isopropanol. Films were drop casted from solutions described above and annealed in the glovebox on a hot plate at 50 °C for 30 min, followed by 80 °C for 30 min. The thickness of the film was ~7.0 μm and the weight of the film was ~ 0.15-0.2 mg.

Structure characterization

The GIWAXS measurement was carried out on a Xeuss 2.0 system (Xenocs, France), which is equipped with a MetalJet source (Excillum, Sweden) (λ = 1.34144 Å). The scattering signal was collected by a Pilatus3R 1M detector. An incidence angle of 0.2° was selected, which was well above the critical angle of typical organic materials, hence probing the inner film structures. The distance between the samples and the detector was 218.349 mm, and the exposure time was 1800 s. The crystallinity of the samples were characterized using X-ray diffraction (XRD, Rigaku SmartLab) with Cu Ka (k = 0.15418 nm) radiation. The thickness of the films was characterized by a Dektak 150 surface profiler. The characterization of the nanostructures was performed using a field emission scanning electron microscope (ZEISS-Merlin, Oberkochen, Germany) at room temperature. The UV-vis-NIR characterization was performed on a SHIMADZU UV-3600 spectrophotometer (Kyoto, Japan). EPR spectra were recorded on a Bruker E500 EPR spectrometer (300 K, 9.854 GHz, X-band, Karlsruhe, Germany). The microwave power used was 6.325 mW and the magnetic field sweep width ranged from 3267 to 3766 Oe. The modulation frequency was 100 kHz and the modulation amplitude was 5 Oe. Cyclic Voltammograms (CV) were acquired using a three-electrode system (CHI600E, Shanghai Chenhua). External beam PIXE experiments were carried out at the GIC4117 1.7-MV tandem accelerator at Beijing Normal University.

Physical properties measurement

The magnetization measurements were performed using Quantum Design PPMS-9 with a vibrating sample magnetometer in the temperature range of 2-400 K. The diamagnetic correction was performed using a diamagnetic susceptibility from the sample holder, silicon substrate and respective Pascal constants. Resistance and Hall resistivity were measured by a standard four-probe method by Quantum Design PPMS-9.

Beamline configuration

XANES spectra at the carbon K-edges were obtained in the BL08U1A beamline at Shanghai Synchrotron Radiation Facility, where polarized X-rays are incident on the sample at an incidence angle of 15°, 60°. The spectra were measured under a magnetic field of 2 kOe at 300 K. In the measurements, total electron yield (TEY) mode is chosen, which usually collects the signal from the topmost 5-10 nm of the sample. The XMCD signal was obtained as the difference between the averages of 10 spectra acquired in the X-ray absorption spectra recorded with right- and left-circularly polarized beam.

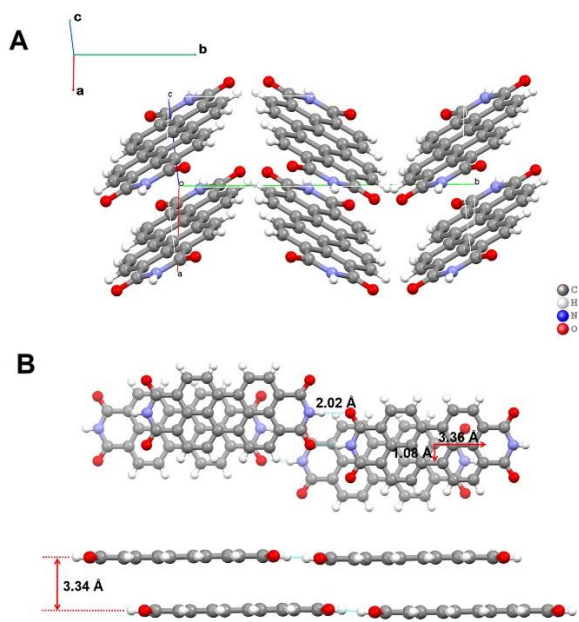

**Fig. S1. Single crystal structure of the sublimated PDI crystallites.** (**A**) Molecular stacking modes of PDI in the single crystal structure. (**B**) The corresponding π-π stacking spacing is 3.34 Å.

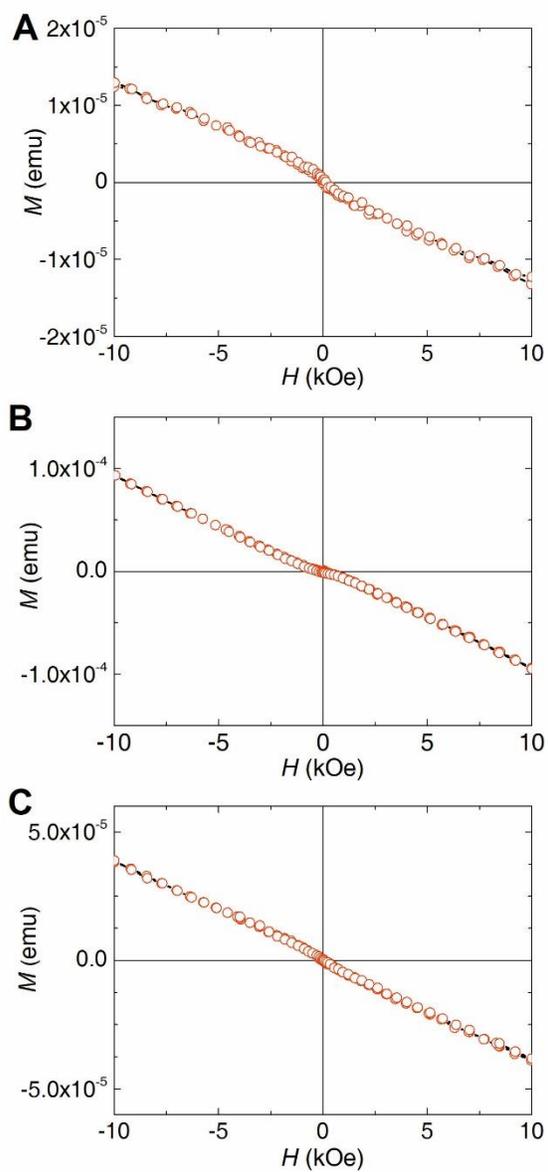

**Fig. S2. Diamagnetism of silicon substrate and PDI raw material.** (**A**), (**B**), and (**C**) Linear *M-H* curves recorded at 300 K indicate a diamagnetic background of silicon substrate, PDI raw material and as-prepared PDI film, respectively.

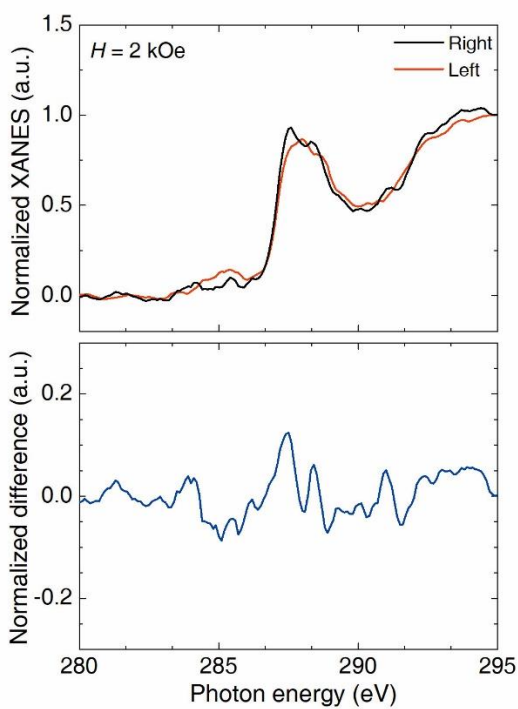

**Fig. S3. X-ray absorption and magnetic dichroic spectra of PDI film at an incidence angle of 60°.** The intensity ratios for the π*-derived resonances peaks show a distinct changes with incidence angle and XMCD signals, defined as the difference in absorption between right- and left-circularly polarized X-rays, show two prominent peaks around 287.4 and 288 eV, which indicates molecular plane is nearly normal to the substrate surface.

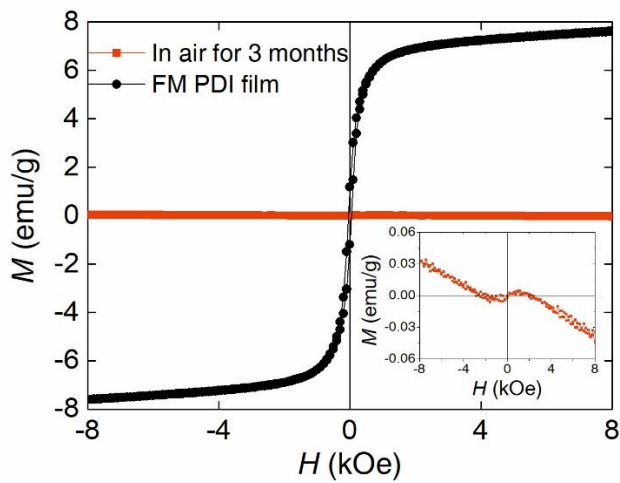

**Fig. S4. Evolution of magnetic properties for ferromagnetic PDI film with time in ambient air.** Insert shows the diamagnetic behavior for PDI film after keeping in ambient air for 3 months.

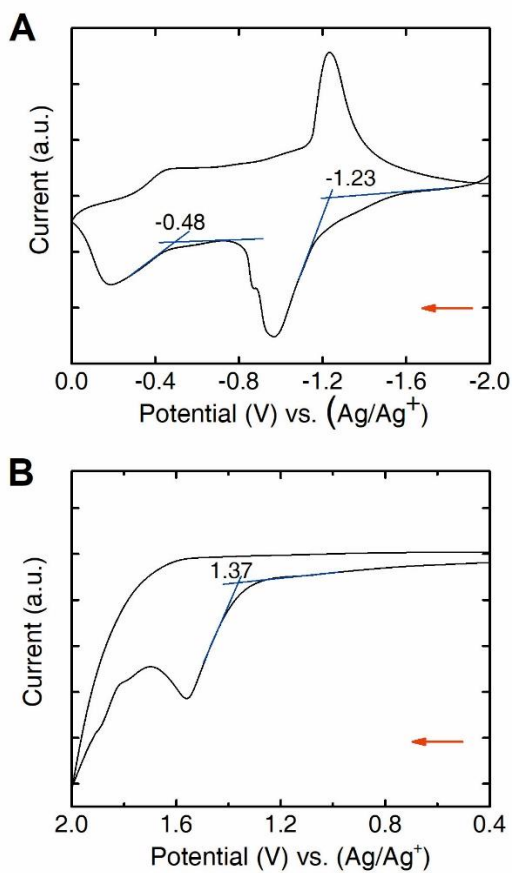

**Fig. S5. CV curves of the PBI dianion film in ACN.** (**A**) From -2.0 V to 0 V. (**B**) From 0.4 V to 2.0 V. The film for measurements was used without exposure to ambient atmosphere. Reference electrode: Ag/Ag$^+$, working electrode: glassy carbon electrode, auxiliary electrode: Pt disk. 0.1 M Bu$_4$NPF$_6$, Fc/Fc$^+$, 298 K, ACN, scan rate: 100 mV s$^{-1}$. The CV was conducted in an inert environment. The PDI$^{2-}$ solution was drop cast on the glassy carbon electrode to form films. The highest occupied molecular orbital (HOMO) level of PDI$^{2-}$, singly occupied molecular orbital (SOMO) level of PDI$^{•-}$ and the HOMO of PDI neutral state could be calculated by equations (1), (2) and (3), respectively. The reference electrode (Ag/Ag$^+$) was calibrated by ferrocene/ferrocenium (Fc/Fc$^+$) as the redox couple.

$$E_{HOMO}^{dianion} = -(4.8 + E_{OX}^{dianion} - E_{1/2}^{Fc/Fc^+}) = -(4.8 - 1.23 - 0.086) = -3.48\ V \quad (1)$$

$$E_{SOMO}^{radical} = -(4.8 + E_{OX}^{radical} - E_{1/2}^{Fc/Fc^+}) = -(4.8 - 0.48 - 0.086) = -4.23\ V \quad (2)$$

$$E_{HOMO}^{neutral} = -(4.8 + E_{OX}^{neutral} - E_{1/2}^{Fc/Fc^+}) = -(4.8 + 1.37 - 0.086) = -6.08\ V \quad (3)$$

$E_{1/2}^{Fc/Fc^+}$ is the half-wave potential of ferrocene/ferrocenium (0.086 V in this work).

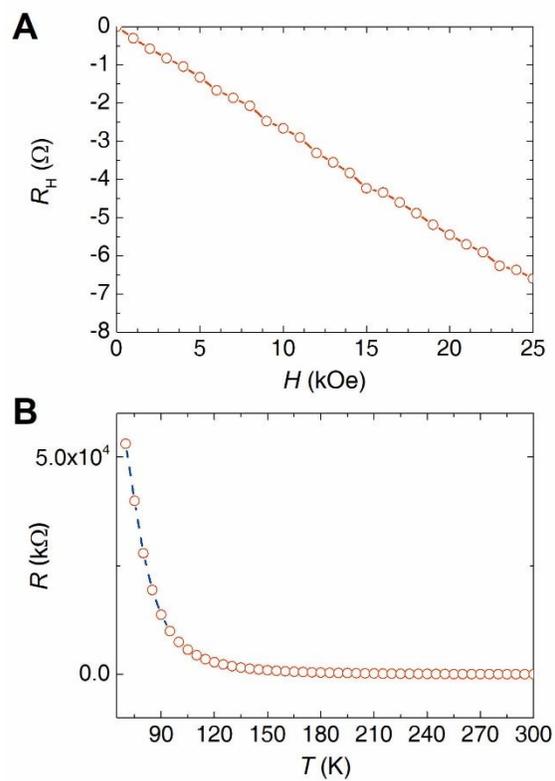

**Fig. S6. Semiconducting properties of the PDI film.** (**A**) Magnetic field dependence of Hall resistance ($R_H$) of the PDI film at 300 K. (**B**) Temperature dependence of resistance of the PDI film. The results show a typical n-type semiconducting feature.

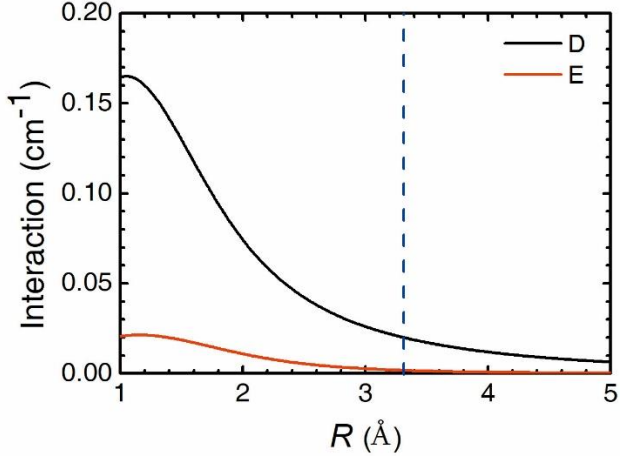

**Fig. S7. Spin interactions between two radicals.** We consider a lattice composed of carbon atoms, and two nearest-neighbor radicals with spin-half form a spin one. The Hamiltonian of spin interactions in the hydrocarbons can be commonly written as $H_S = D\hat{S}_z^2 + E(\hat{S}_x^2 - \hat{S}_y^2)$, where $D$ and $E$ denote the interaction strengths with respect to the molecular field, and $\hat{S}$ the spin operator. The parameters are $D = g^2 \mu_B^2 \lambda / 4(1-M^2)$, $E = g^2 \mu_B^2 \gamma / 4(1-M^2)$ with $g$ factor being 2 and $\mu_B$ the Bohr magneton. Using the Gaussian-type orbitals $x\exp(-\alpha R^2)$ with $x$ being $R\sqrt{2\alpha}$, $\lambda$, $\gamma$ and $M$ have been derived as *(27)*,

$$\lambda = \frac{1}{2R^3}\left[P(x)(135/x^4 - 24/x^2 + 4) - 2Q(x)(135/x^3 + 21/x + 5x + 107x^3/105)\right],$$

$$\gamma = \frac{3}{R^3}\left[P(x)(-15/x^4 + 2/x^2) + 2Q(x)(15/x^3 + 3/x + x/3 + x^3/105)\right],$$

and $M = \exp(-x^2/4)$, where $P(x) = \frac{1}{\sqrt{2\pi}}\int_{-x}^{x}\exp(-t^2/2)dt$, $Q(x) = \frac{1}{\sqrt{2\pi}}\exp(-x^2/2)$. For carbons, $\alpha$ is determined to be 1.44 Å$^{-2}$ by maximizing the overlap between Gaussians and relevant Slater functions. The curves of $D$ and $E$ for various $R$ are thus calculated as displayed here. In our cases, the π-π distance is 3.32 Å as indicated by the blue dashed line, so the spin interaction is 0.02 cm$^{-1}$, on the same order of the coercive field observed in our experiment.

**Table S1. Metal impurities elemental analysis by PIXE for the ferromagnetic PDI film.** The saturation magnetization contributions of these metal impurities to the sample is no more than $5\times10^{-3}$ emu·g$^{-1}$, which is far smaller than the measured ferromagnetic results in our experiments.

| Mn (ppm) | Fe (ppm) | Co (ppm) | Ni (ppm) |
|---|---|---|---|
| N/A | 16 | 0.8 | N/A |